\documentclass[preprints,article,accept,moreauthors,pdftex]{Definitions/mdpi} 

\firstpage{1} 
\pubvolume{1}
\issuenum{1}
\articlenumber{0}
\pubyear{2021}
\copyrightyear{2020}
\datereceived{} 
\dateaccepted{} 
\datepublished{} 
\hreflink{https://doi.org/} 
\Title{Searching for high-$z$ radio galaxies with the MGCLS}

\TitleCitation{Searching for high-$z$ galaxies with the MGCLS}

\Author{Kenda Knowles$^{1,2,\dagger*}$\orcidA{}, Sinah M. Manaka$^{1,\dagger}$\orcidB{},  Michael F. Bietenholz$^{3,4}$\orcidC{}, William D. Cotton$^{5,2}$\orcidD{}, Matthew Hilton$^{6,7}$, Konstantinos Kolokythas$^{8}$, S. Ilani Loubser$^{8}$\orcidG{}, and Nadeem Oozeer$^{2,9}$\orcidH{} }

\AuthorNames{Kenda Knowles, Sinah M. Manaka, Michael F. Bietenholz, William D. Cotton, Matthew Hilton, Konstantinos Kolokythas, S. Ilani Loubser, and Nadeem Oozeer...}

\AuthorCitation{Knowles, K.; Manaka, S.M.; Bietenholz, M.F.; Cotton, W.D.; Matthew Hilton; Kolokythas, K.; Loubser, S.I.; Oozeer, N.}

\address{%
$^{1}$ \quad Centre for Radio Astronomy Techniques and Technologies, Department of Physics and Electronics, Rhodes University, Makhanda, South Africa\\
$^{2}$ \quad South African Radio Astronomy Observatory (SARAO), 2 Fir Street, Observatory, South Africa\\
$^{3}$ \quad Hartebeesthoek Radio Astronomy Observatory, SARAO, Krugersdorp 1740, South Africa\\
$^{4}$ \quad York University, Toronto, ON, Canada\\
$^{5}$ \quad National Radio Astronomy Observatory, Charlottesville, VA 22903, USA\\ 
$^{6}$ \quad School of Mathematics, Statistics \& Computer Science, University of KwaZulu-Natal, Westville Campus, Durban 4000, South Africa\\ 
$^{7}$ \quad Astrophysics Research Centre, University of KwaZulu-Natal, Westville Campus, Durban 4000, South Africa\\ 
$^{8}$ \quad Centre for Space Research, North-West University, Potchefstroom, 2520, South Africa\\
$^{9}$ \quad African Institute for Mathematical Sciences, 6-8 Melrose Road, Muizenberg 7945, South Africa\\
}
\corres{Correspondence: k.knowles@ru.ac.za}

\firstnote{These authors contributed equally to this work.}

\abstract{We present results from a search for high-redshift radio galaxy (H$z$RG) candidates using 1.28\,GHz data in the Abell~2751 field drawn from the MeerKAT Galaxy Cluster Legacy Survey (MGCLS). We use the H$z$RG criteria that a radio source is undetected in all-sky optical and infrared catalogues, and has a very steep radio spectrum. We use the likelihood ratio method for cross-matching the radio catalogue against multi-wavelength galaxy catalogues from the Dark Energy Camera Legacy Survey (DECaLS) and the All-sky Wide Infrared Survey Explorer (AllWISE). For those radio sources with no multi-wavelength counterpart, we further implement a radio spectral index criterium of $\alpha < -1$, using in-band spectral index measurements from the wide-band MeerKAT data. Using a 5$\sigma$ signal-to-noise cut on the radio flux densities, we find a total of 274 H$z$RG candidates: 179 ultra-steep spectrum sources, and 95 potential candidates which cannot be ruled out as they have no spectral information available. The spectral index assignments in this work are complete above a flux density of 0.3\,mJy, at least an order of magnitude lower than existing studies in this frequency range or when extrapolating from lower frequency limits. Our faintest H$z$RG candidates with and without an in-band spectral index measurement have a 1.28\,GHz flux density of 57\,$\pm$\,8\,$\mu$Jy and 68\,$\pm$\,13\,$\mu$Jy, respectively. Although our study is not complete down to these flux densities, our results indicate that the sensitivity and bandwidth of the MGCLS data makes them a powerful radio resource to search for H$z$RG candidates in the Southern sky, with 20 of the MGCLS pointings having similar image quality as the Abell~2751 field and full coverage in both DECaLS and AllWISE. Data at additional radio frequencies will be needed for the faintest source populations, which could be provided in the near future by the MeerKAT UHF band (580 -- 1015 MHz) at a similar resolution ($\sim$\,8-10$^{\prime\prime}$).}

\keyword{high-redshift galaxies; radio continuum} 

\begin{document}

\section{Introduction}

Observing the high redshift Universe is critical to our understanding of cosmological and astrophysical processes. It allows us to probe large-scale structure, galaxy evolution, and the period of time when the first stars formed, known as the epoch of reionisation ($z > 6$). In the radio regime, the high-redshift Universe can be probed through high redshift radio galaxies (H$z$RGs). These galaxies, with large volumes of dust and gas \citep[e.g.,][]{1996ApJ...460...68E, 2005A&A...430L...1D}, are expected to be actively forming stars and may be the progenitors of the massive ellipticals seen in the local Universe \citep{1998MNRAS.295..549B, 2001MNRAS.323..417A}. With their link to high density environments such as proto-clusters or the centre of galaxy clusters (see the review by \citet{2008A&ARv..15...67M} and references therein), H$z$RGs are critical targets to understand the formation and evolution of galaxies at early times, and their connection to large scale structure formation.

Creating complete samples of H$z$RGs is challenging. The radio regime does however produce optically unbiased samples as it is unaffected by dust absorption, and is therefore the most efficient band in which to detect high redshift galaxies \citep{2010MNRAS.405..436I}. \citet{1982PASP...94..397S} was the first to identify a radio galaxy beyond $z \sim 1$, with the highest redshift radio galaxy currently known being at $z = 5.72$ \citep{2018MNRAS.480.2733S}. Current searches for H$z$RG candidates require an additional data source, either in the form of multi-wavelength catalogues, or multi-frequency radio data. 

As H$z$RGs are by definition at high redshift ($z \gtrsim 2$), the effect of redshift dimming can be exploited with H$z$RG candidates identified as lacking a multi-wavelength counterpart when cross-matching with optical and/or infrared (IR) sky survey catalogues \citep{2012MNRAS.420.2644K}. Where near-IR data is available, the $K-z$ relation \citep{1984MNRAS.211..833L} has been used to identify potential H$z$RGs when an infrared counterpart is found \citep[e.g.,][]{2001MNRAS.326.1585J,2006ApJ...649...63S}. \citet{2006AJ....132.2409N} discovered a population of infrared-faint radio sources (IFRSs) after cross-matching radio and IR data for one of the Australia Telescope Large Area Survey fields. IFRSs are characterised by a radio source with a very faint or non-detection in deep IR data and may be a related to, or a subset of, H$z$RGs as they are also found at high redshifts \citep{2011A&A...526A...8M,2014MNRAS.439..545C,2014A&A...567A.104H}.

Where multi-frequency radio data are available, H$z$RGs candidates have been identified through megahertz-peaked spectrum (MPS) sources \citep{2004NewAR..48.1157F,2015MNRAS.450.1477C}, or by exploiting the now established correlation between galaxy redshift and radio spectral steepness\footnote{Here we assume the flux density convention $S_\nu \propto \nu^\alpha$ where $\alpha$ is the spectral index.} \citep[$z - \alpha$;][]{1994A&AS..108...79R, 2010MNRAS.405..436I}. The latter method uses ultra-steep spectrum (USS; $\alpha < -1$)\footnote{Many earlier studies use a USS criterium of $\alpha < -1.3$.} radio sources to isolate H$z$RGs candidates. The correlation is not well-understood, with several theories put forward as to what causes the spectral steepening at high redshift \citep[e.g.,][]{1991AJ....102.1659K, 2006MNRAS.371..852K, 2010MNRAS.406..197B, 2012MNRAS.420.2644K, 2018MNRAS.480.2726M}, and the correlation is not exact as USS sources do not represent the entire high-$z$ population \citep[e.g.,][]{2003NewA....8..805P, 2009MNRAS.398L..83J}. However, implementing the USS criteria has yielded the largest numbers of H$z$RG candidates \citep{2000A&AS..143..303D,2006MNRAS.373.1531C,2007MNRAS.381..341B,2011ApJ...743..122A,2011JApA...32..609I,2014A&A...569A..52S,2014ApJ...793...82V,2018MNRAS.475.5041S} and, when combined with multi-wavelength cross-matching, can produce robust candidate lists for spectroscopic follow-up \citep{2003NewAR..47..303R,2007MNRAS.378..551B,2009MNRAS.395.1099B,2010ApJ...725...36D, 2011BASI...39..539G,2019MNRAS.489.5053S}.

Early searches for H$z$RG candidates were restricted to bright ($S_{1.4\,GHz} > 1 - 10$\,mJy) radio sources extracted from available wide-area radio surveys \citep[e.g.,][]{2002A&A...394...59D,2009MNRAS.395.1099B}. This flux density limit is gradually being reduced thanks to sensitive imaging with the new generation of radio interferometers \citep[see e.g.,][]{2018MNRAS.475.5041S}, although implementation of the USS cut typically still requires multi-frequency radio observations to obtain spectral indices. The wideband receivers of the current pre-SKA era telescopes may be able to circumvent this requirement. The advancements in the quality of radio data, coupled with the availability of wide-area optical and infrared surveys, such as the Dark Energy Camera Legacy Survey \citep[DECaLS;][]{2019AJ....157..168D} and the All-sky Wide-field Infrared Survey Explorer \citep[All-WISE;][]{2014yCat.2328....0C}, are making H$z$RG searches more accessible. 


In this paper we present the first results of a search for H$z$RG candidates in the Abell~2751 field using deep, wideband imaging from the MeerKAT Galaxy Cluster Legacy Survey \citep[MGCLS;][]{MGCLS}. We cross-match the 1.28\,GHz data with the DECaLS and AllWISE catalogues, and use spectral index information from the wide MeerKAT band to implement USS cuts. The paper is structured as follows. In Section~\ref{sec:data} we introduce the radio data used in this study. In Section~\ref{sec:method} we describe the multi-stage methodology used to determine H$z$RG candidates. The results are discussed in Section~\ref{sec:discussion} with a summary and concluding remarks given in Section~\ref{sec:conclusion}. Throughout this article we assume a flat $\Lambda$CDM cosmology with $H_0 = 70$\,km/s/Mpc and $\Omega_m = 0.3$.
 
\section{Radio catalogue: MGCLS Abell 2751 field} \label{sec:data}


We use the first data release (DR1) of the compact source catalogue from the MeerKAT Galaxy Cluster Legacy Survey \citep[MGCLS;][]{MGCLS}. The MGCLS consists of long-track (6--10 hour) observations of 115 galaxy clusters in the Southern sky using MeerKAT's L-band receiver (900 -- 1670\,MHz). For each cluster field, a primary beam-corrected image is produced at an effective observing frequency of 1.28\,GHz. Each primary beam-corrected image spans a 1.2$^\circ\,\times$\,1.2$^\circ$ sky region with a typical central RMS noise of 4$-$6\,$\mu$Jy\,beam$^{-1}$ in the full-resolution survey ($\approx$\,7.5$^{\prime\prime}-$8$^{\prime\prime}$ full-width half-maximum (FWHM)). The source catalogue for each field was generated using the Python Blob Detection and Source Finder (\textsc{pybdsf}) package \citep{2015ascl.soft02007M} with the default thresholds (3$\sigma$ island RMS and 5$\sigma$ peak RMS), and the final compact source catalogue includes only those sources identified as having only a single Gaussian component by \textsc{pybdsf}. Note that the dimensions of the elliptical Gaussian were free in \textsc{pybdsf}, so single-component sources are not necessarily unresolved.

The MGCLS DR1 images have not been calibrated for direction-dependent effects and therefore some fields present artefacts around the brightest sources. The level of contamination varies with the source brightness and position of the source within the primary beam. The \textsc{pybdsf} source finder incorrectly identifies some of these artefacts as real sources. To mitigate the level of artefact contamination in the final DR1 compact source catalogue, all catalogue sources within a given radius of a sufficiently bright source are removed. With this method, some real sources are removed along with the contaminating artefacts. However, the total number of sources removed (real and artefact) amounts to $\sim$\,2\% of the total number of catalogued sources, and the removal of real sources is not expected to have a large effect on this work. 

For this study we use the Abell~2751 cluster field, centred at RA\,$\rm=\,00^h 16^m 13.92^s$, Dec.$\,=\,-31^\circ 23^\prime 18.60^{\prime\prime}$, with coverage from the DECaLS and AllWISE catalogues (see Section~\ref{sec:mwcats}). The primary beam-corrected image has low RMS noise (6.6--11\,$\mu$Jy\,beam$^{-1}$ at $\sim$\,7.8$^{\prime\prime}$ resolution)\footnote{The noise increases from the centre of the image outwards due to primary beam-correction.} and minimal visual contamination by bright source artefacts. The artefact-excised DR1 compact source catalogue for the Abell~2751 field consists of 3610 radio sources.

\section{Identifying H\emph{z}RG candidates} \label{sec:method}

We use a multistage approach to identify possible H$z$RG candidates in our radio catalogue, using sequential cuts to arrive at the final list of candidates. Given the number of sources involved, we use in-house automated tools where possible. In this section we discuss each step of our process and the resultant whittling down of the catalogue to the final candidate list. Table~\ref{tab:dropout} indicates the fraction of potential candidates remaining after each step of the process. 
\begin{specialtable}[H] 
\caption{Selection criteria\label{tab:dropout}}
\begin{tabular}{llcc}
\toprule
\textbf{Step No.} & \textbf{Selection} & \textbf{Number of Sources} & \textbf{Fraction}\\
\midrule
0  & Compact MGCLS sources   & 3610  & 100\% \\
1a & 5$\sigma$ SNR limit           & 1700  & 47\% \\
1b & $<\,10^{\prime\prime}$ angular size limit    & 1414  & 39\% \\
2a & Cross-match: DECaLS     & 587 & 16\%  \\
2b & Cross-match: AllWISE    & 415 & 11\%  \\
3  & $\alpha < -1.0$\quad$(\alpha_{\rm none})$ &  231 (109)  & 6 (3) \%  \\
4  & Manual cross-checks    &   179 (95)  & 5 ($\sim$3) \%  \\ 
\bottomrule
\end{tabular}
\end{specialtable}
\subsection{Step 1: Signal-to-noise and angular size requirements}

Many H$z$RG searches enforce a flux density cut, with the most recent studies using sub-mJy limits at low (sub-GHz) frequencies \citep[e.g.,][]{2018MNRAS.475.5041S}. Given the sensitivity of the MGCLS data, we could potentially probe down to $S_{\rm 1.28\,GHz} \sim 35$\,mJy at the 5$\sigma$ level. However, although we selected one of the MGCLS fields with the fewest bright-source artefacts, the image noise does vary across the field due to the primary beam correction, with the local RMS noise increasing away from the image centre. Rather than implement a fixed flux density cut, we apply a signal-to-noise ratio (SNR) restriction using the fitted flux density versus its associated uncertainty from the source catalogue\footnote{\textsc{pybdsf} calculates uncertainties as per \citet{1997PASP..109..166C}.} as the first step of the process. After applying a fitted flux density SNR requirement of 5$\sigma$, the radio catalogue is reduced to 1700 sources. {It is possible that the input catalogue discussed in Section~\ref{sec:data} may be missing faint sources due to the source finding parameters. However, as the faintest sources are excluded through this SNR cut, this is not expected to affect the number of potential H$z$RG candidates in this study.}

We simultaneously apply an angular size restriction of $s_{\rm maj} \leq 10^{\prime\prime}$. Angular size restrictions have been necessary in H$z$RG searches using lower resolution (lower frequency) data in order to preferentially select high redshift sources \citep{2018MNRAS.475.5041S}. Our radio data have a resolution of $\sim$\,7.8$^{\prime\prime}$ so this requirement may not be necessary given that we are using a compact/point source catalogue. However, investigation of the catalogue shows that several sources, although meeting the single Gaussian fit requirement, are well resolved. Following \citet{2018MNRAS.475.5041S}, we therefore restrict the catalogue to sources with an angular size smaller than $\sim 10^{\prime\prime}$. There are 1414 radio sources after applying the SNR and angular size limits.

\subsection{Step 2: Multi-wavelength cross-matching}\label{sec:mwcats}

The next step of the process is to determine which radio sources have neither optical nor infrared counterparts using publicly available catalogues. To discard relatively bright nearby galaxies, we first cross-match our SNR-cut catalogue against the eighth data release of the Dark Energy Camera Legacy Survey \citep[DECaLS;][]{2019AJ....157..168D}. The DECaLS catalogue includes \emph{g}-, \emph{r}-, and \emph{z}-band magnitudes, and has a positional accuracy of $\approx$\,20\,mas. The sources with no identified DECaLS counterpart are then cross-matched against the All-sky Wide-field Infrared Survey Explorer (AllWISE) catalogue \citep{2014yCat.2328....0C} which provides mid-infrared source data. We use the AllWISE W1, W2, and W3 band data, which have 5$\sigma$ median sensitivity limits of 54\,$\mu$Jy, 71\,$\mu$Jy, and 730\,$\mu$Jy respectively, and astrometric accuracies in the Abell~2751 region of better than 70\,mas.

We use the likelihood ratio (LR) method, introduced by \citet{1992MNRAS.259..413S} and further adapted by \citet{2003A&A...398..901C}, to cross-match our radio and multi-wavelength catalogues and determine optical or infrared counterparts for our radio sources. Here we summarise the key properties of the LR method and refer the reader to  \citet{1992MNRAS.259..413S} and \citet{2009MNRAS.397..623G} for more details. An LR value is determined for each potential match within a specified search radius of a given radio source, where the LR compares the probability that a source match is the true counterpart to that of the same source being a spurious match, i.e.,

\begin{equation}
    \mathrm{LR} = \frac{q(m)f(r)}{n(m)},
\end{equation}

where $q(m)$ is the expected magnitude distribution for the true counterparts, $f(r)$ is the probability distribution function of the positional uncertainties in the involved catalogues, and $n(m)$ is the surface density of the unrelated background objects with magnitude \emph{m}. For our catalogues, the probability distribution function is a two-dimensional Gaussian distribution dependent on the combined positional uncertainties of the sources in both catalogues. We assume elliptical Gaussian distributions for all catalogues: we use radio positional uncertainties from the MGCLS catalogue, IR positional uncertainties from the AllWISE catalogue, and a conservative systematic positional uncertainty of 0.2$^{\prime\prime}$ in both RA and Dec. for DECaLS.

In cases where there is more than one potential match to a given radio source, the LR method can be used to calculate the reliability, $R_i$, of each match:

\begin{equation}
    R_i = \frac{{\rm LR}_i}{{\sum\rm LR_{radius}} + (1-Q)}\quad,
\end{equation}

where $\sum$LR$_{\rm radius}$ is the sum of LR for all possible counterparts to a radio source within our defined search radius, and $Q$ is the fraction of the radio sources with catalogue counterparts above the magnitude limit. The LR method can therefore be used to identify the \emph{most likely} counterpart for a given source in cases where there may be more than one possible match. An estimate of the spurious identification rate (or error rate) can be obtained by comparing $R_i$ with the total number of counterparts above a certain LR value. This cutoff in LR aims to optimize the completeness versus error rate of the cross-matching, where completeness is defined as the fraction of the radio catalogue that is assigned a counterpart. The LR method implementation in the \textsc{astromatch}\footnote{\url{https://github.com/ruizca/astromatch}} code used in this work provides a facility to automatically optimize the LR cutoff and we use the code-generated optimal cut in our cross-matching here. 

\begin{figure}
\widefigure
\includegraphics[width=0.46\textwidth,clip=True,trim=30 0 0 30]{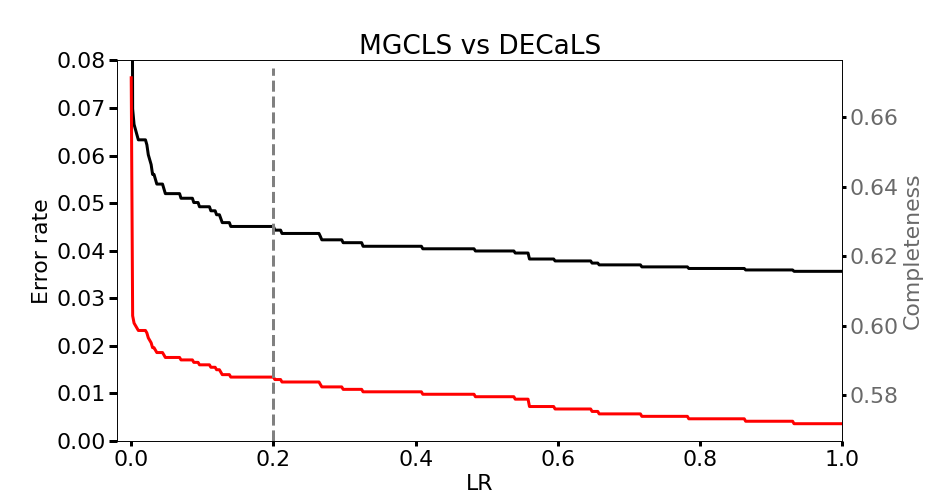}\hspace{0.5cm}
\includegraphics[width=0.46\textwidth,clip=True,trim=30 0 0 30]{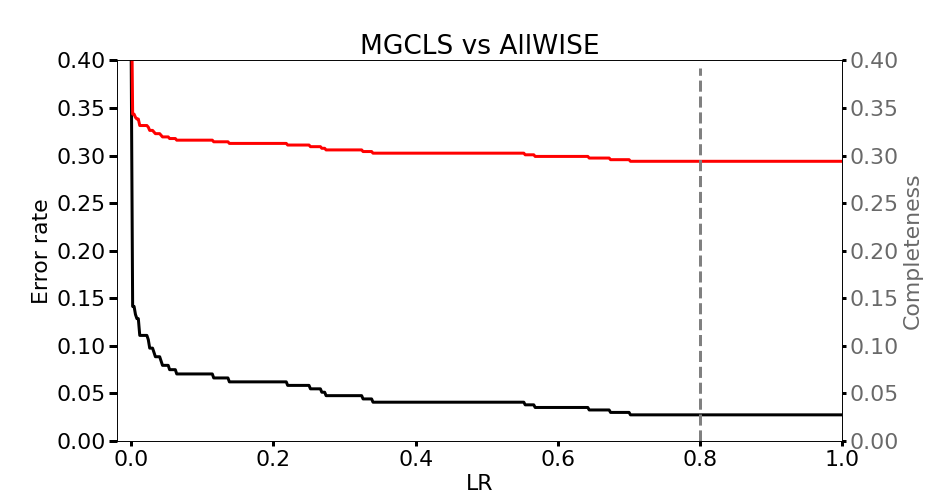}
\caption{Cross-matching error rate (black) and completeness (red) versus LR. \textit{Left:} MGCLS--DECaLS cross-match results. The vertical dashed line indicates the optimal threshold LR > 0.2.\label{fig:LRdecals} \textit{Right:} MGCLS--AllWISE cross-match results where the vertical dashed line indicates the optimal LR cutoff of 0.8.\label{fig:LRallwise} }
\end{figure}

We use a search radius of 4$^{\prime\prime}$ for the MGCLS--DECaLS and MGCLS--AllWISE cross-matching, optimised to produce the smallest number of spurious versus real matches. At this search radius, the optimal cutoff in LR is 0.2, shown by the vertical dashed line in the left panel of Figure~\ref{fig:LRdecals}. With LR > 0.2 our cross-matching has a completeness of 59\% with an error rate less than 4\%. Considering the best matches only, 587 of the radio sources have no identified optical counterpart. We cross-match these against the AllWISE catalogue. The right panel of Figure~\ref{fig:LRdecals} shows the completeness and error rate versus all values of LR for the MGCLS--AllWISE cross-matching. The optimal LR cutoff is 0.8 in this case, with $\sim$\,29\% completeness and an error rate less than 3\%. We are left with 415 radio sources with no optical or mid-infrared counterpart.


\end{paracol}
\begin{paracol}{2}
\switchcolumn


\subsection{Step 3: USS selection}\label{sec:method-uss}
There is coverage of this field at 843\,MHz from the Sydney University Molonglo Sky Survey \citep[SUMSS;][]{2003MNRAS.342.1117M}, however, the SUMSS data are of significantly lower resolution than the MGCLS data and source confusion would prohibit accurate spectral index determinations for all but the most isolated sources. There is also coverage from the Very Large Array Sky Survey \citep[VLASS;][]{2020PASP..132c5001L} which provides $\sim$\,2.5$^{\prime\prime}$-resolution data at 3\,GHz. Cross-matching the list of candidates from the previous step, we obtain $\alpha^{\rm 3 GHz}_{\rm 1.28 GHz}$ spectral indices for only 13 sources. We therefore exploit the wideband nature of the MGCLS data, which covers a $\sim$\,700\,MHz range in frequency (fractional bandwidth of 0.55), to extract in-band spectral index values for the remaining radio sources with no identified multi-wavelength counterpart.

One of the MGCLS DR1 products is a 12-plane frequency cube for each field. The effective GHz observing frequency of each plane is 0.908, 0.952, 0.997, 1.043, 1.093, 1.145, 1.317, 1.381, 1.448, 1.520, 1.594, and 1.656, respectively. We perform \textsc{pybdsf} source finding, using default settings, in each plane and cross-match the resultant source catalogues to build up the flux density distribution for each source. We use a weighted Levenberg-Marquardt least-squares fitting procedure to fit a power-law spectrum of the form $S_\nu \propto \nu^\alpha$ to each distribution. We only produce a fit if a source is detected in at least two frequency planes -- if the source has only two flux density measurements, we further require a minimum frequency separation of 200\,MHz in order to proceed with the fit. {As flux density measurements are taken from the same dataset, with sub-band images having identical resolution, any spatial variability in the flux density scale should affect all sub-band images. For a given source, we therefore do not expect any bias in the spectral index fitting due to this effect.}

We obtain a spectral index fit for 306/415 sources from step~2. Implementing a USS cut of $\alpha < -1$ (suitable for low flux density sources; see \citep{2014A&A...569A..52S} for details), we have 231/415 potential H$z$RG candidates. There are 109/415 radio sources which do not have spectral index information as they did not have sufficient flux density information for a fit. We retain these sources as potential candidates.

\subsection{Step 4: Manual cross-checking with NED and Vizier}

As a final step, we manually check the 340 ($231 + 109$) candidates against the NASA/IPAC Extragalactic Database \citep[NED;][]{1991ASSL..171...89H} to look for any redshift information, or whether a source exists in other optical/IR catalogues. We use a NED search radius of 4$^{\prime\prime}$. Six of the H$z$RG candidates have available redshift information, five of which have $z < 0.12$ and are therefore discarded from the candidate list. The sixth source, MKTCS~J001802.40$-$313505.5, is confirmed as a high-redshift galaxy at $z = 2.13$ \citep[H-ATLAS\,J001802.2$-$3135005;][]{2018MNRAS.473.1751B}. An additional 60 sources have NED cross-matches from other optical/IR catalogues and are discarded as candidates. Finally, we use Vizier \citep{2000A&AS..143...23O} to cross-match against The Million Quasars catalogue \citep[v7.2,][]{2021yCat.7290....0F} -- we find no matches within 4$^{\prime\prime}$ of our candidates. Our final H$z$RG candidate list contains a total of 274 sources: 179 USS sources and 95 potential H$z$RG candidates with no spectral index information. 

\end{paracol}
\begin{specialtable} 
\caption{Excerpt of the final catalogue of 274 H$z$RG candidates after implementation of an SNR and angular size cut, DECaLS and AllWISE catalogue cross-matching, USS selection, and manual checks with NED. 25\% of the candidates have no spectral index information and are retained as possible H$z$RGs (see Section~\ref{sec:method-uss} for details). The catalogue includes the IAU MGCLS source identifier, the J2000 radio position in decimal degrees, the 1.28\,GHz flux density and associated statistical uncertainty, the source size (major and minor axes, and p.a.), and the fitted in-band spectral index (where available). Note that the sizes are after convolution with the CLEAN beam, and the intrinsic sizes will typically be smaller. The full catalogue is available online at \url{https://github.com/kendak333/MGCLS_HzRGs}. \label{tab:candidates}}
\begin{tabular}{lcccccccc}
\toprule
\textbf{Source ID}	& \textbf{RA}	& \textbf{Dec.} & $\mathbf{S_{\mathbf\mathrm 1.28 GHz}}$ & $\mathbf{{\Delta}S_{\mathbf\mathrm 1.28 GHz}}$ & $\mathbf{s_{\mathbf\mathrm maj}}$ & $\mathbf{s_{\mathbf\mathrm min}}$ & $\mathbf{s_{\mathbf\rm p.a.}}$ & ${\alpha}$ \\
	& \textbf{(deg)}	& \textbf{(deg)} & \textbf{(mJy)} & \textbf{(mJy)} & \textbf{(arcsec)} & \textbf{(arcsec)} & \textbf{(deg)} &  \\
\midrule
MKTCS J001349.00-314001.7 & 3.4542 & -31.6672 & 0.351 & 0.025 & 8.4    & 7.9    & 120     & -1.86 \\
MKTCS J001349.01-305328.5 & 3.4542 & -30.8913 & 0.295 & 0.039 & 9.5    & 7.8    & 118     & -2.65 \\
MKTCS J001349.46-314127.3 & 3.4561 & -31.6909 & 0.168 & 0.033 & 9.6    & 9.2    & 127     &       \\
MKTCS J001350.21-305923.4 & 3.4592 & -30.9898 & 0.268 & 0.031 & 8.5    & 7.7    & 123     & -1.32 \\
MKTCS J001351.03-315758.6 & 3.4627 & -31.9663 & 0.475 & 0.042 & 9.0    & 8.2    & 128     & -2.63 \\
MKTCS J001351.62-311725.8 & 3.4651 & -31.2905 & 0.157 & 0.025 & 9.0    & 8.1    & 105     & -3.25 \\
MKTCS J001358.51-310818.0 & 3.4938 & -31.1384 & 0.215 & 0.031 & 10.0   & 8.1    & 75      & -2.71 \\
MKTCS J001359.59-312348.2 & 3.4983 & -31.3967 & 0.140 & 0.025 & 9.5    & 8.7    & 59      & -2.09 \\
MKTCS J001359.64-314617.2 & 3.4985 & -31.7715 & 0.172 & 0.030 & 9.1    & 8.1    & 77      &       \\
MKTCS J001359.80-305256.1 & 3.4992 & -30.8823 & 3.288 & 0.032 & 7.9    & 7.4    & 96      & -1.07 \\
\bottomrule
\end{tabular}
\end{specialtable}

\begin{figure}
\widefigure
\includegraphics[height=6.3cm]{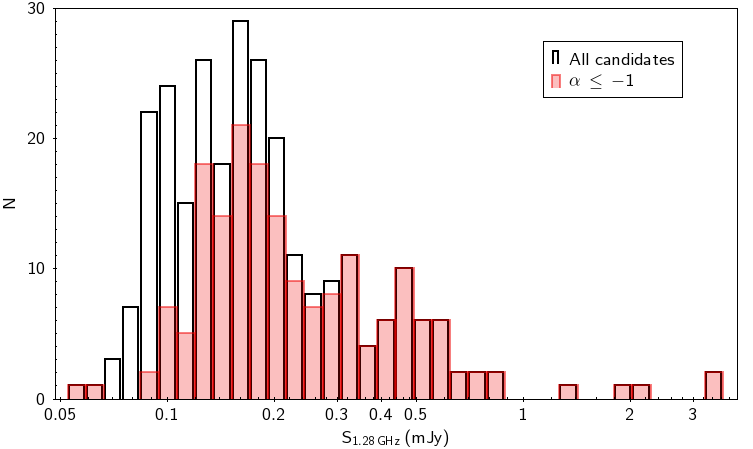}\hspace{0.2cm}
\includegraphics[height=6.3cm]{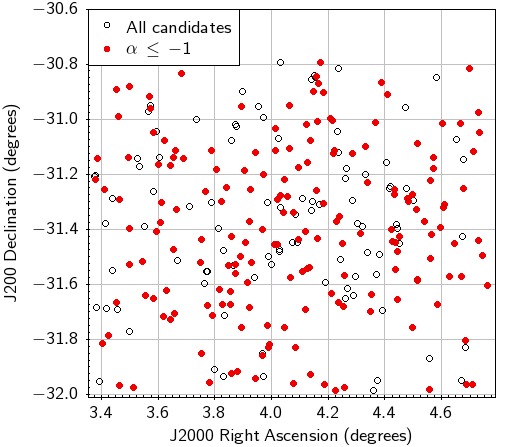}
\caption{\textit{Left:} Flux density distribution of the 274 H$z$RG candidates (empty bins). The distribution of the 179 USS sources is shown by the shaded bins. Spectral index measurements for candidates are complete above 0.3\,mJy, and largely incomplete below $\sim$\,120\,$\mu$Jy. \textit{Right:} Sky distribution of the H$z$RG candidates, with the USS sources by filled red circles. Clusters of candidates may indicate positions of potential proto-clusters. 
\label{fig:distributions} }
\end{figure}
\begin{paracol}{2}
\switchcolumn

\section{Discussion} \label{sec:discussion}
An excerpt of the final list of 274 H$z$RG candidates is given in Table~\ref{tab:candidates}. The candidates amount to 8\% of the sources in the Abell~2751 MGCLS compact source catalogue, with spectroscopic follow-up or high quality photometry required to confirm the redshift of the sources. 

The flux density distribution of all the candidates is shown in the left panel of Figure~\ref{fig:distributions}. As expected, a large fraction of the fainter sources ($S_{\rm 1.28\,GHz} \lesssim 120\,\mu$Jy) are lacking a spectral index measurement due to the source having insufficient SNR across the band. However, we have complete spectral index coverage above a flux density of 0.3\,mJy, which is the faintest source population inspected for H$z$RG candidates at GHz frequencies to date. 

To investigate a potential positional effect (e.g., from the primary beam correction at different frequencies) on the ability to assign a spectral index value for a source, we plot the sky positions of all 340 candidates in Figure~\ref{fig:distributions}. The sources are coloured by whether the source has a spectral index or not, and we see no positional trend with respect to the undetermined $\alpha$ values. This may be a positive indication that the SNR cut adequately takes into account the changes in noise levels across the primary beam-corrected field of view for these data.


\section{Summary and Conclusion} \label{sec:conclusion}
We have presented the results of a search for high-redshift radio galaxy (H$z$RG) candidates at 1.28\,GHz using one of the 115 cluster fields, namely Abell~2751, from the MeerKAT Galaxy Cluster Legacy Survey \citep[MGCLS;][]{MGCLS}. We used a multi-stage approach, first implementing a $5\sigma$ signal-to-noise and $< 10^{\prime\prime}$ angular size cut on the radio catalogue before cross-matching against catalogues from DECaLS and AllWISE. Taking advantage of the wideband MeerKAT data, we were able to determine in-band spectral index measurements for 74\% of the radio sources which had no multi-wavelength counterpart after the cross-matching. Using an ultra-steep spectrum cut of $\alpha < -1$, and performing manual checks for supplementary redshift information in NED, we compiled a final catalogue of 274 H$z$RG candidates, 95 of which have no available spectral index information and cannot be excluded. Our spectral index coverage is complete above a flux density of 300\,$\mu$Jy. This is the lowest flux density at GHz frequencies to be probed for H$z$RGs to date.

This study shows the great potential in the MGCLS DR1 data, and indeed any long-track MeerKAT L-band data, for finding H$z$RG candidates. With 20 of the 115 MGCLS fields having good dynamic range and full coverage in both DECaLS and AllWISE, extending this study to these fields may produce more than $\sim$ 4000 H$z$RG candidates for spectroscopic follow-up. Additional frequency radio data will still be needed in order to obtain spectral index measurements for sources in the sub-100\,$\mu$Jy flux density ranges probed by MeerKAT. MeerKAT's UHF band, with $\sim\,7.5^{\prime\prime}$ resolution at near-uniform weighting, could provide such data with a small request of telescope time.








\vspace{6pt} 



\authorcontributions{K.K. and S.M. contributed equally to this work with their main responsibilities as follows: K.K. -- conceptualization, interpretation of results, analysis, draft writing and editing; S.M. -- implementing methodology, analysis, draft writing. Other contributions: draft review -- M.F.B., W.C., M.H., K.Ko., I.L., N.O.; data resources -- W.C., M.H.; methodology (cross-matching) -- M.H., I.L., N.O. All authors have read and agreed to the submitted version of the manuscript.}

\funding{K.Kn. is supported by the New Scientific Frontiers grant of the South African Radio Astronomy Observatory. S.M. acknowledges funding from the South African Radio Astronomy Observatory and the National Research Foundation (NRF Grant Number: 97800). I.L. is partially funded by National Research Foundation of South Africa (NRF Grant Number: 120850). }

\dataavailability{The MGCLS data products used in this work are proprietary until the survey paper \citep{MGCLS} has been accepted for publication. }

\acknowledgments{MGCLS data products were provided by the South African Radio Astronomy Observatory and the MGCLS team and were derived from observations with the MeerKAT radio telescope. The MeerKAT telescope is operated by the South African Radio Astronomy Observatory, which is a facility of the National Research Foundation, an agency of the Department of Science and Innovation. DECam data were obtained through the Legacy Surveys imaging of the DESI footprint, which is supported by the Director, Office of Science, Office of High Energy Physics of the U.S. Department of Energy under Contract No. DE-AC02-05CH1123, by the National Energy Research Scientific Computing Center, a DOE Office of Science User Facility under the same contract; and by the U.S. National Science Foun-
dation, Division of Astronomical Sciences under Contract No. AST-0950945 to NOAO. DECam data were also obtained from the Astro Data Lab at NSF’s National Optical-Infrared Astronomy Research Laboratory. NOIRLab is operated by the Association of Universities for Research in Astronomy (AURA), Inc. under a cooperative agreement with the National Science Foundation. The NASA/IPAC Extragalactic Database is operated by the Jet Propulsion Laboratory, California Institute of Technology, under contract with the National Aeronautics and Space Administration. This research has made use of the VizieR catalogue access tool, CDS, Strasbourg, France (DOI: \url{10.26093/cds/vizier}). The original description of the VizieR service was published in A\&AS 143, 23.}

\conflictsofinterest{The authors declare no conflict of interest.}

\end{paracol}



\reftitle{References}
\externalbibliography{yes}
\bibliography{mgcls-hzrgs}

\end{document}